\documentclass[aps,amsmath,amssymb]{revtex4}

\usepackage{amsmath}    
\usepackage{amssymb}    
\usepackage{amsfonts}   
\usepackage{graphicx}
\begin{document}

\author{F. Stauffer$^{1}$}
\affiliation{$^1$ Laboratoire de Physique Th\'eorique, Louis Pasteur University,
Strasbourg, France}

\date{\today}

\title{Two-level relationships and Scale-Free Networks}


\begin{abstract}
Through the distinction between ``real'' and ``virtual'' links between the nodes of a graph, we develop a set of simple rules leading to scale-free networks with a tunable degree distribution exponent. Albeit sharing some similarities with preferential attachment, our procedure is both faster than a na\"ive implementation of the Barab\'asi and Albert model and exhibits different clustering properties. The model is thoroughly studied numerically and suggests that reducing the set of partners a node can connect to is important in seizing the diversity of scale-free structures.
\end{abstract}
\maketitle

\section{Introduction}

In the last decade, the study of complex networks sharing similarities with structures like the World Wide Web has drastically developed. Their broad interdisciplinary scope led to a large numbers of models aimed at explaining why they are not falling in the random graphs category studied by Erd\"os and R\'enyi in the late fifties \cite{erdos59}.
A quite exhaustive review on the matter can be found in \cite{bararmp}. The now famous concept of preferential attachment \cite{barascience} succeded in explaining why the degree distribution $P(k)$  -- the probability of finding a node with $k$ nearest neighbors-- of such heterogeneous networks decays algebraically, which echoes the empirical results obtained from various data. Usually the exponent $P(k)\propto k^{-\gamma}$ ranges between 2 and 3, with some exceptions such as SPIRES co-authorship \cite{newman} ($\gamma = 1.2$) and sexual contacts \cite{liljeros} ($\gamma = 3.4$). Several modifications to this rule (aging \cite{aging}, or initial attractiveness \cite{initalattrac} for instance) enlarged the range of predicted exponents from the initial $\gamma=3$ calculated by Barab\'asi and Albert.

Though it has been observed that (almost) linear preferential attachment occurs in real networks \cite{pareal}, it suffers from several drawbacks. Besides the limited insight on the microscopic details, it is rather slow and not convenient to generate large scale free structures from a numerical point of view. This is a consequence of the unrestricted partner choice given to each newly added node. In the same spirit, it does not take in account other constraints existing in real graphs as for instance in \cite{barrat}, and focuses only on their topology. Respecting the needed balance between attractiveness and randomness can be achieved opting for a different road \cite{vazquez,dorogovtsev}.

To speed up the generation of large scale free structures, one can for instance look at the algorithm recently developed by Mukherjee and Manna \cite{muker} or earlier by Krapivski and Redner in \cite{krap1,krap2}. 
The former is a blatantly fast algorithm but it is more or less equivalent to the Barab\'asi and Albert (B-A) model as far as the rest of the properties of the generated structures are concerned. 

The aim of this paper is to propose a simple model achieving the previous improvements without relying on the B-A model, opening up the possibility for different topological properties (such as peculiar clustering) to emerge. The plan of the paper is the following: we will first introduce the algorithm we constructed. We will then proceed to the numerical study of the model, discuss the degree distributions obtained, the clustering properties and the average distance between nodes on the network.

\section{Model and Implementation}

One of the simplest algorithm generating scale-free networks with $\gamma=3$ is detailed in Ref.  \cite{dorogovtsev}. Based on connecting newly added nodes to both ends of a randomly chosen link on the network, it takes a constant time to add a vertex throughout the growth process. Looking at the situation from a social point of view, we know that in general people often meet through the help of some ``third-party'' person. This idea meets to some extent the process of selecting a random link on the network. If we consider the problem of a physicist asking another physicist advice to find a new collaborator, a two-level relationship can appear. Some people might forward to someone he already worked with (a real connection) or to someone he just heard of or met but with whom he has no direct relationship with (a virtual connection or ``acquaintance''). In general, this two-level hierarchy in relations is relevant and embodies the main ingredient we will rely upon. In our network, two nodes will share a real link whenever they are physically connected. A node $i$ will share a virtual link with node $j$ either because they  are  already connected by a real link or because they have an indirect relation.

Let us take a look at the details of our model. The nodes of our network will each maintain two sets: the set $\mathcal{L}$ representing all the real links it makes and the list of it's virtual links $\mathcal{V}$ without excluding the case where $\mathcal{V} \cap \mathcal{L} \neq  \emptyset$. Our model is based on the linear growth of the network using the following rules:
\begin{enumerate}
\item At every simulation step, a new node $i$ is added to the network.
\item It chooses a random partner $j$ and, with a probability $Q$, a second random partner $j'$ among the virtual links of $j$.\label{fs1}
\item $i$ adds $j$ and $j'$ (if it exists) in among it's real connections, but \emph{only} pushes $j'$ back in his virtual links.\label{fs2}
\item $j$ and $j'$ add $i$ in both their sets  $\mathcal{L}$ and $\mathcal{V}$.
\item $j'$ eventually spreads it's ``popularity'' to $R$ sites on the network\footnote{avoiding itself}, from nearest neighbor to nearest neighbor, by adding itself to their list $\mathcal{V}$.\label{fs3}
\end{enumerate}

\begin{figure}
\centerline{\includegraphics[scale=0.50]{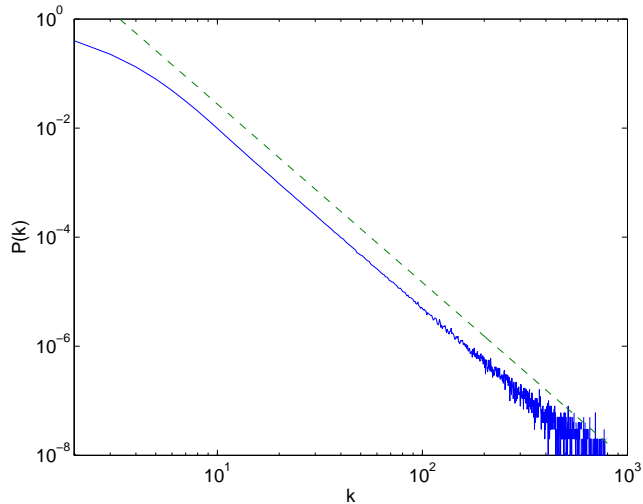}}
\caption{Degree Distribution $P(k)$ in  double logarithmic scale, for $N=10^6$ sites with $Q=1$ and $R=0$. The dashed line is a power-law fit with $\gamma= 3.3$.}
\label{fr1}
\end{figure}

Up to step (\ref{fs1}), the procedure looks  similar to \cite{dorogovtsev}, but after (\ref{fs2}) the discrepancy introduced between the sets  $\mathcal{V}$ and $\mathcal{L}$ changes the situation. In the case of  multiple encounters, even if a real connections occurs between all the nodes involved, $i$ \emph{only remembers about the last node it connected to}. Then, this ``memory'' is consolidated in the long-term, and $i$ remembers about his newly acquired acquaintance $j'$ forever. The last two steps are extra-rules aimed at gaining control on $\gamma$. Spreading the popularity represents the external influences helping nodes to find out who are the most connected amongst them, and it also leads to potential redundancy in the set $\mathcal{V}$.  Taking $Q<1$ corresponds to the introduction of ``socially inept'' nodes who are reluctant to make new connections.

We shall introduce several notations: let $\langle N(k)\rangle$ be the average number of sites with $k$ real links in the stationary regime.
We also define the average number of virtual links of degree $q$  for nodes of degree $k$ $\langle n(q)\rangle_k$ and $\langle n(q)\rangle=\sum_k \langle n(q)\rangle_k$. Obtaining a truly useful analytical description of our model is challenging, and we decided to go for a numerical simulation to study the behavior of the system.

Before we describe our results, we recall that the clustering of a node $i$ is defined by the following relation,
\begin{equation}
C_i = 2\frac{e_i}{k_i(k_i-1)},
\end{equation}
where $k_i$ is the degree of the node, and $e_i$ the number links between the $k_i$ neighbors of $i$. $C_i=1$ is the largest value reached by the clustering as  $k_i(k_i-1)/2$ is the maximum number of undirected links between $k_i$ nodes. For $R=0$, our procedure always connects a node to two partners who are already connected, greatly enhancing the clustering. We then have $k_i-1$ links between the neighbors of a site of a degree $k_i$. We get the following exact analytical expression for the clustering in this case:
\begin{equation}\label{eq.cluster}
C_i = \frac{2}{k_i}.
\end{equation}
which was used as a safety check to test our code.
\section{Numerical Results}

We shall start by inspecting the effect of the existence of real and virtual links only. For that matter, we set $R=0.0$ and $Q=1.0$. The statistical data gathered was obtained on networks with at least $10^6$ sites. A single run takes only several seconds on a 1.8 GHz PPC 970 based desktop computer with 1.25 Gb of RAM. The memory overhead resulting of having to maintain two sets for each site becomes quite steep. Our implementation was written in C++ which slightly increased the memory requirements as well. We emphasize on the fact that C++ remains an excellent language for this kind of simulation.

Each simulation consisted in several realizations (typically 100) of networks with constant size $N$. We computed the degree distribution $P(k)$, the cumulative degree distribution $P_{CDF}(k) = P(X>k)$, the average clustering, the clustering per site and the average distance between two nodes on the network. To illustrate the competition between real and virtual links, we also plotted the degree distribution of all the virtual links on the network 
$n(k) = \sum_q \langle n(k)\rangle_q$, which grows like $3t$. 
We also measured the individual values of $\langle n(k)\rangle_q$ for different values of $q$.  Up to a very good approximation the degree distribution follows a power law with exponent $\gamma\approx 3.3\pm 0.1$, for over 3 decades (see figure \ref{fr1}). 

\begin{figure}
	\begin{tabular}{lclc}
		(a) & \includegraphics[scale=0.3]{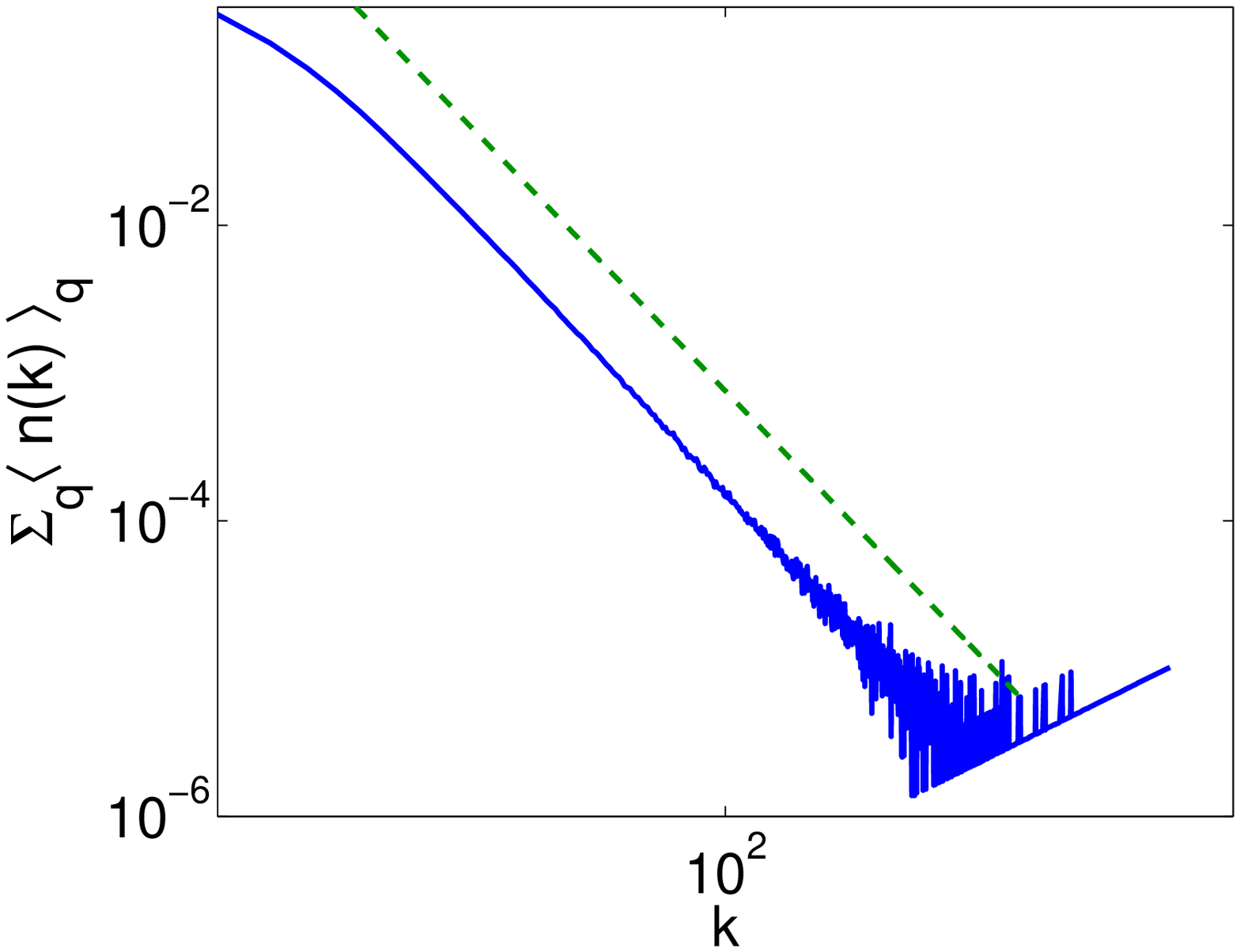} & (b) & \includegraphics[scale=0.3]{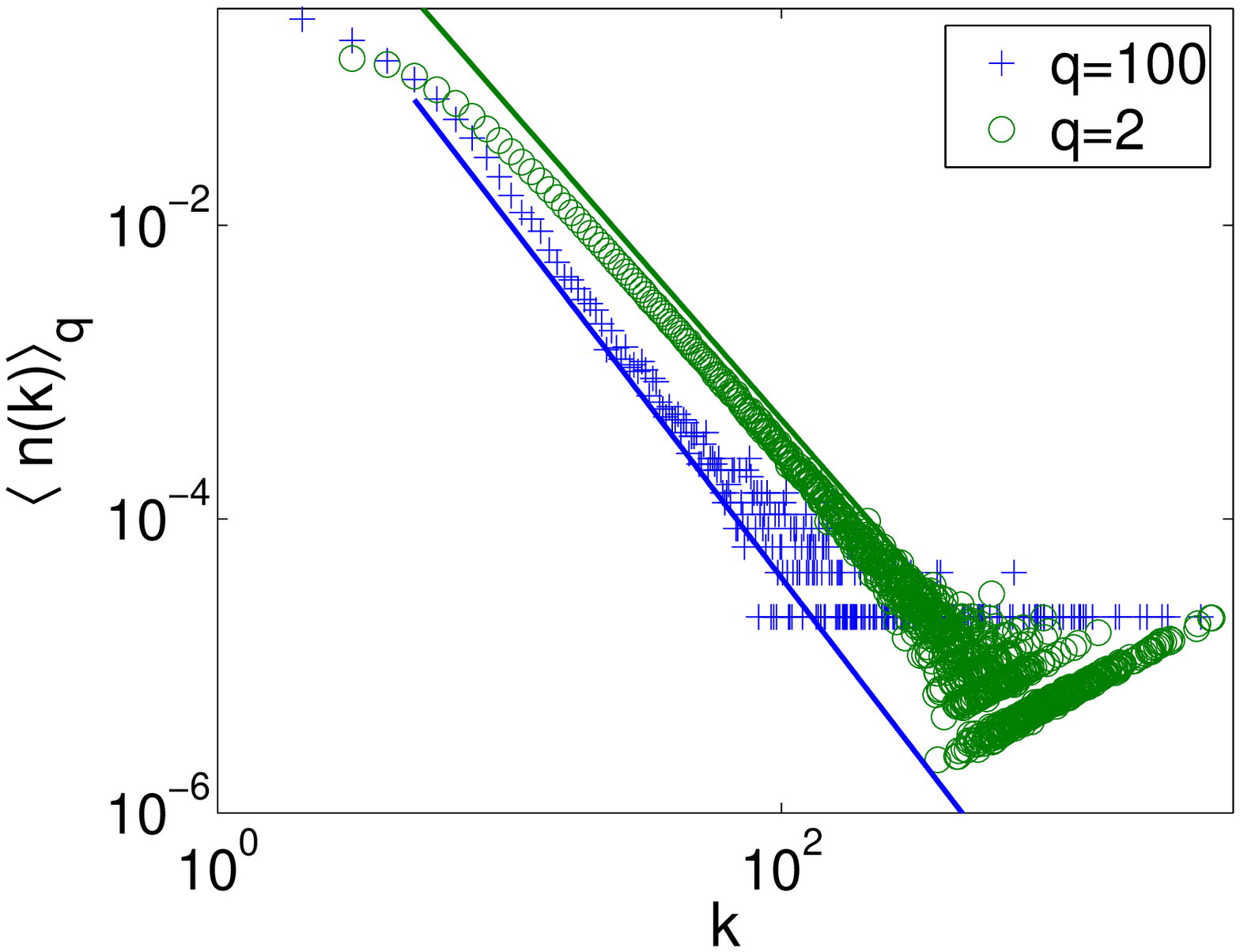} 
	\end{tabular}
\caption{ (a): Total degree distribution of virtual links averaged on 100 realizations of networks with $10^6$ nodes ($Q=1,~R=0$) with power law fit where $\gamma=2.1$. (b): Individual virtual link distribution for $q=2,100$ and corresponding fits with $\gamma=2.2, 2.5$ respectively. }
\label{fr6}
\end{figure}

As one can expect the tail's exponent is larger than $3$, because the ``short-term memory'' in case of multiple encounters leads to missing several potential interesting partners with a large degree. We emphasize on the fact the growth quickly stabilizes to the power-law degree distribution. It seems that our procedure biases the growth enough to speed up this process. The average clustering is quite high, $<C>\approx 0.69$, as it is mostly dominated by the effect of sites having exactly two neighbors (with $C_i=1$). Interestingly, it doesn't seem to decay with $N$ unlike the B-A model. The average distance between two sites on the network is about 14 sites with a standard deviation of 3 for $N=10^6$ nodes. The most interesting point is that the total degree distribution of virtual links $\sum_q \langle n(k)\rangle_q$ seems to have no typical scale as well (see figure \ref{fr6}-(a)) , and that it's tail exponent is close to $\gamma\approx 2.1$. There is one notable exception from this behavior: for very large values of $k$ there is a crossover between power-law decay and linear growth. Linear growth is natural for large values of $k$ because such nodes most likely make connections to the added sites through a ``third-party'' random node. So that they will end up in close to $k$ different virtual links list, \textit{i.e.}:
\begin{equation}
\sum_q \langle n(k)\rangle_q \propto k ~~~\text{when } k\gg 1
\end{equation}


Should this crossover occur earlier, we would have lost the power-law scaling by making it to easy for the rich to get richer. The individual distribution $\langle n(k)\rangle_q$ also exhibits a scale-free behavior like the total distribution, except we only have a linear tail for small values of the degree $q$. For instance, we measured $\gamma\approx 2.2$ for $q=2$ and $\gamma\approx 2.5$ for $q=100$, the general aspect of the distributions is given in figure \ref{fr6}-(b).  This means that on the average, the least connected sites share mostly relations with the most connected ones, whereas the most connected ones do not have particular relations between them. Virtual links fasten the process of building the largely connected nodes by keeping track of them. 

\begin{figure}
\centerline{\includegraphics[scale=0.5]{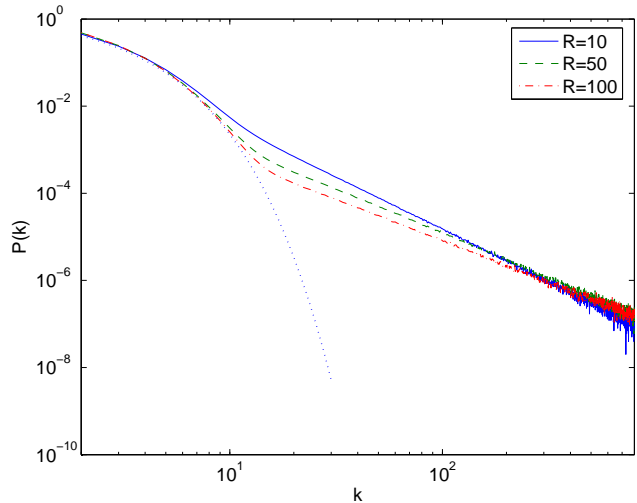}}
\caption{Logarithm of the Degree Distribution versus the logarithm of the degree for networks with $N=10^6$ sites and $Q=1.0$ for different values of R. Power-law tails are given $\gamma = 2.4, 2.0, 1.9 \pm 0.1$ for $R=10,50,100$ respectively. The dotted line represents an exponential decay illustrating visually the random nature of the graph for low $k$. }
\label{fr3}
\end{figure}

\begin{figure}
\centerline{\includegraphics[scale=0.5]{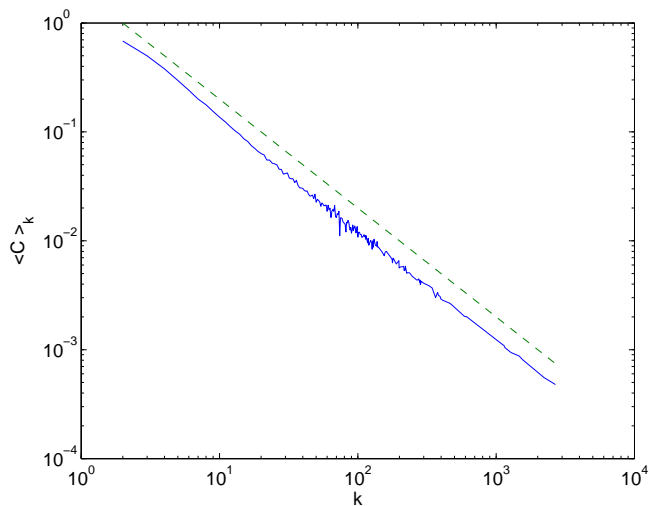}}
\caption{Average clustering versus degree. The full line corresponds to data obtained for $Q=1.0$ and $R=10$. The dashed line to Eq. (\ref{eq.cluster}).}
\label{fr7}
\end{figure}

Increasing the value of $R$ does not destroy the scale-free behavior. Numerical estimates for the exponents,  $\langle C \rangle$ and $\langle l \rangle$ are given in Table \ref{t.1}. 
Figure (\ref{fr3}) illustrates those distributions. We see that for  low values of the degree $k$, the distribution is exponential, but after a decade in the log-log scale the power-law tail builds up. The exponential start, enhanced visually by the decrease of the exponent $\gamma$ simply illustrates the fact that the choice of the first partner is fully random, hence we can expect the nodes mostly involved in the random selection process (for which $P(k)\lessapprox 10^{-3}$ ) will form an homogeneous network. Memory requirement were too high to go beyond the limit $R=100$ and $N=10^6$. Clustering decreases on the average, but we observed very small deviations in the power-law $\langle C \rangle_k$ suggesting that it remains proportional to $k ^{-1}$ up to a very good precision (see figure (\ref{fr7})). It is worthwhile noting that our propagation procedure strongly interplays with the clustering. The average distance between two nodes decreases as well, along with it's standard deviation. The virtual links distributions do not chance much from $R=0$. We end by exploring the $Q < 1, R=0$ where  observed a slow increase of the exponent $\gamma$ beyond $3.3$ until numerical simulations become pointless due to the unavoidable finite size effects. 

\begin{table}
\begin{center}
\begin{tabular}{|c|c||c|c | c | }
\hline $Q$  & $R$ & $\gamma$ & $<C>$ & $<l>$\\
\hline\hline  1 & 0 & $3.3\pm 0.1$ & 0.69 & $14 \pm 3$ \\
\hline   & 10 &$2.4\pm 0.1$ & 0.55 & $7 \pm 2$    \\
\hline   & 50 & $2.0 \pm 0.1$ & 0.26 & $5 \pm 1$ \\
 \hline  & 100 & $1.9 \pm 0.1$ & 0.16 & $5 \pm 1$ \\
 \hline
\end{tabular}
\end{center}
\caption{Exponents, average clustering and average value of the shortest path for different values of $R$ ($Q=1$).}
\label{t.1}
\end{table}

\section{Summary and Conclusion}

We studied numerically a model which reproduces the scale-free behavior observed in heterogeneous graphs such as the World Wide Web. Reducing the set of potential partners newly created nodes can connect to by introducing a two-level relationship between the nodes. The competition between the two relations is highly non-trivial, and it has been observed that both distribution of ``real'' and ``virtual'' links are scale-free, albeit linear tails are often observed in the latter.

Adding two extra ingredients: the possibility for a node to spread it's popularity on the network, and the presence of ``socially inept'' sites helped us to increase the range of exponents observed until finite size effect become too important to rely on computer simulation. The networks obtained are quite clustered independently of their size, and clustering seems to decay like $\frac{1}{k}$ in any case we studied. The implementation is fast enough to allow the efficient growth of very large network, achieving our initial objective. The indirect reduction of the potential list of partners for a newly created node introduced by our procedure can be linked to other direct methods aimed at speeding up preferential attachment. 


The author would like to thank R. M., R. D.,  A. T. and J. P. for useful discussions and encouragements.





\appendix

\end{document}